\documentclass[twocolumn]{bmcart}
\usepackage[utf8]{inputenc} 

\usepackage{hyperref}
\usepackage{amsmath}
\usepackage{algorithm}
\usepackage{algorithmicx}
\usepackage[noend]{algpseudocode}
\usepackage{epsfig}
\usepackage{cite}
\usepackage{subfigure}
\usepackage{color}
\usepackage{multirow}
\usepackage{makecell}
\usepackage{sidecap}
\usepackage{afterpage}
\usepackage{pdflscape}

\begin{document}
\begin{frontmatter}
\begin{fmbox}

\title{Fast and Scalable Inference of Multi-Sample Cancer Lineages}

\author[
]{\inits{VP}\fnm{Victoria} \snm{Popic}$^1$}
\author[
]{\inits{RS}\fnm{Raheleh} \snm{Salari}$^1$}
\author[
]{\inits{IH}\fnm{Iman} \snm{Hajirasouliha}$^1$}
\author[
]{\inits{DK}\fnm{Dorna} \snm{Kashef-Haghighi}$^1$}
\author[
]{\inits{RW}\fnm{Robert B} \snm{West}$^2$}
\author[
   corref={aff1},            
   email={serafim@cs.stanford.edu}  
]{\inits{SB}\fnm{Serafim} \snm{Batzoglou}$^1$}


\vspace{15pt}

\begin{center} 
{\sffamily$^1$ Department of Computer Science, Stanford University, Stanford, CA}\\
{\sffamily$^2$ Department of Pathology, Stanford University School of Medicine, Stanford, CA}
\end{center}

\begin{abstract} 
Somatic variants can be used as lineage markers for the phylogenetic reconstruction of cancer evolution. Since somatic phylogenetics is complicated by sample heterogeneity, novel specialized tree-building methods are required for cancer phylogeny reconstruction. We present LICHeE ($L$ineage $I$nference for $C$ancer $H$eterogeneity and $E$volution), a novel method that automates the phylogenetic inference of cancer progression from multiple somatic samples. LICHeE uses variant allele frequencies of SSNVs obtained by deep sequencing to reconstruct multi-sample cell lineage trees and infer the subclonal composition of the samples. LICHeE is open-sourced and available at \url{http://viq854.github.io/lichee}.
\end{abstract}

\end{fmbox}

\end{frontmatter}

\section*{Background}
Cancer is driven by the accumulation of somatic mutations that confer fitness advantages to the tumor cells. Numerous studies have shown tumors to be highly heterogeneous, consisting of mixtures of cell subpopulations with distinct sets of somatic variants (for example see review papers \cite{1,Marusyk2012}). With the advent of next-generation sequencing technologies, many large-scale efforts are underway to catalog the somatic mutational events driving the progression of cancer  \cite{2, 3} and infer the phylogenetic relationships of tumor subclones. Characterizing the heterogeneity and inferring tumor phylogenies are key steps for developing targeted cancer therapies \cite{Fisher2013} and understanding the biology and progression of cancer. 

In order to reconstruct tumor phylogenies, studies have utilized variant allele frequency (VAF) data of somatic single nucleotide variants (SSNVs) obtained by whole-genome  \cite{4, 5}, exome  \cite{6}, and targeted deep sequencing  \cite{4, campbell2008subclonal}. Clustering of SSNVs based on VAF similarity  \cite{nik2012life,shah2012clonal,roth2014pyclone} and detection of copy number aberrations, while accounting for variable sample purity \cite{6, oesper2013inferring, ha2014titan}, have been used to differentiate and order groups of mutational events. While many evolutionary studies of cancer have focused on single-sample intra-tumor heterogeneity \cite{hajirasouliha2014combinatorial}, several studies have also compared multiple tumor samples extracted from a single patient either at different points in time during cancer progression  \cite{ding2012clonal, landau2013evolution, mcfadden2014genetic} or from different regions of the same tumor or its metastases  \cite{5, campbell2010patterns,yachida2010distant,15,green2013hierarchy, de2014spatial}. In multi-sample approaches, the patterns of SSNV sharing (i.e., distinguishing somatic mutations that are omnipresent, partially shared, or private among the samples) can serve as phylogenetic markers from which lineage trees are reconstructed \cite{hajirasouliha2014reconstructing}. On the basis of the lineage trees, the evolutionary timing of each mutational event can then be inferred with high confidence  \cite{5, campbell2010patterns, salari2013inference, landau2013evolution}.

Most existing multi-sample studies with a relatively small number of SSNVs infer the tumor phylogenies manually by analyzing SSNV VAFs and presence patterns across samples  \cite{5, green2013hierarchy, gerlinger2012intratumor}. Several other studies used implementations of traditional phylogeny reconstruction methods, such as neighbor joining with Pearson correlation distances  \cite{19}, or maximum parsimony  \cite{15} on patterns of somatic mutational sharing across samples.  However, in order to scale to datasets comprised of large numbers of samples per patient and extract fine-grained SSNV timing information, as well as handle sample heterogeneity, which traditional tree-building techniques are not designed to do, specialized computational approaches need to be developed for tumor cell lineage reconstruction. Two recent computational methods, {\em SubcloneSeeker} \cite{subcloneseeker} and {\em PhyloSub} \cite{phylosub}, have been developed to address this need. The method {\em SubcloneSeeker} requires as input clusters of variant cell prevalence (CP) estimates that need to be obtained by processing the data using other existing tools. Given the CP clusters, the method generates all possible subclone structures in each sample separately. The per-sample solutions are then trimmed by checking their compatibilities during the merge step. However, the merge step is designed to check compatibilities of two tumor samples only (e.g. relapse/primary tumor sample pairs that are common in clinical studies) and it cannot merge the subclone structures of more than two samples. Currently it only reports which sample trees are compatible across the given pair of samples. The method {\em PhyloSub} performs reasonably on samples with very few mutations that form simple (chain) topologies; however, it produces unsatisfactory results on larger multi-sample datasets, such as \cite{15} (see Appendix A for details).

In this work we introduce LICHeE ($L$ineage $I$nference for $C$ancer $H$eterogeneity and $E$volution), a novel computational method for the reconstruction of multi-sample tumor phylogenies and tumor subclone decomposition from targeted deep sequencing SSNV datasets. Given SSNV VAFs from multiple samples, LICHeE finds the set of lineage trees that are consistent with the SSNV presence patterns and VAFs within each sample and are valid under the cell division process. Given each such tree, LICHeE provides estimates of the subclonal mixtures of the samples by inferring sample heterogeneity simultaneously with phylogenetic cell lineage tree reconstruction. LICHeE is able to search for lineage trees very efficiently by incorporating the SSNVs into an evolutionary constraint network that embeds all such trees and applying VAF constraints to reduce the search space. LICHeE runs in only a few seconds given hundreds of input SSNVs and does not require data preprocessing.

We demonstrate that LICHeE is highly effective in reconstructing the lineage trees and sample heterogeneity by evaluating it on simulated trees of heterogeneous cancer cell lineage evolution, as well as on three recently published ultra-deep sequencing multi-sample datasets of clear cell renal cell carcinoma (ccRCC) by Gerlinger et. al  \cite{15}, high-grade serous ovarian cancer (HGSC) by Bashashati et. al  \cite{19}, and breast cancer xenoengraftment in immunodeficient mice by Eirew et. al \cite{eirew2014dynamics}, for which single-cell validation results are also available. LICHeE found unique trees for each ccRCC patient and for all except one patient (for which multiple valid trees were found) of the HGSC study. For the ccRCC dataset, LICHeE trees were nearly identical to the published trees, which are the result of a multi-step thorough analysis of the data, involving SSNV calling, clustering, and tree-building using maximum parsimony. For the HGSC dataset, LICHeE improved on the results reported by the study, producing trees with better support from the data. LICHeE also revealed additional heterogeneity in the samples of both studies. In particular, LICHeE identified subclones in one more sample of the ccRCC study (in addition to the reported six samples) and three samples of the HGSC study, all supported by the data. Finally, the trees reconstructed by LICHeE on the xenoengraftment dataset can be highly-validated by the single-cell analysis presented in the paper. LICHeE is open-source and freely distributed at  \cite{lichee}, and includes an intuitive graphical user interface (GUI) that may aid users to perform quality control on the output trees as well as interpret the trees biologically. 

\section*{Results and Discussion}
\subsection*{Overview of the Multi-Sample Cancer Phylogeny Inference Method, LICHeE}

LICHeE is a method designed to reconstruct cancer cell lineages using SSNVs from multiple related normal and tumor samples of individual cancer patients, allowing for heterogeneity within each sample. Given a set of validated deeply sequenced SSNVs,  LICHeE uses the presence patterns of SSNVs across samples and their VAFs as lineage markers by relying on the perfect phylogeny model  \cite{gusfield1991efficient}. This model assumes that mutations do not recur independently in different cells; hence, cells sharing the same mutation must have inherited it from a common ancestral cell. This assumption can be used to derive the following SSNV ordering constraints. Firstly, (1) a mutation present in a given set of samples cannot be a successor of a mutation that is present in a smaller subset of these samples, since it could not have arisen independently by chance in the additional samples. Similarly, (2) a given mutation cannot have a VAF higher than that of its predecessor mutation (except due to CNVs), since all cells containing this mutation will also contain the predecessor. Finally, (3) the sum of the VAFs of mutations disjointly present in distinct subclones cannot exceed the VAF of a common predecessor mutation present in these subclones, since the subclones with the descendent mutations must contain the parent mutations (this constraint is formally defined in the Methods section). These constraints provide key information about the topology of the true cell lineage tree and are leveraged by LICHeE to define the search space of the possible underlying lineage trees and evaluate the validity of the resulting topologies. The final goal of LICHeE is to find phylogenetic trees encoding an evolutionary ordering of the input SSNVs that does not violate any of these three constraints.

\begin{figure*}[t!]
\includegraphics[scale=.7]{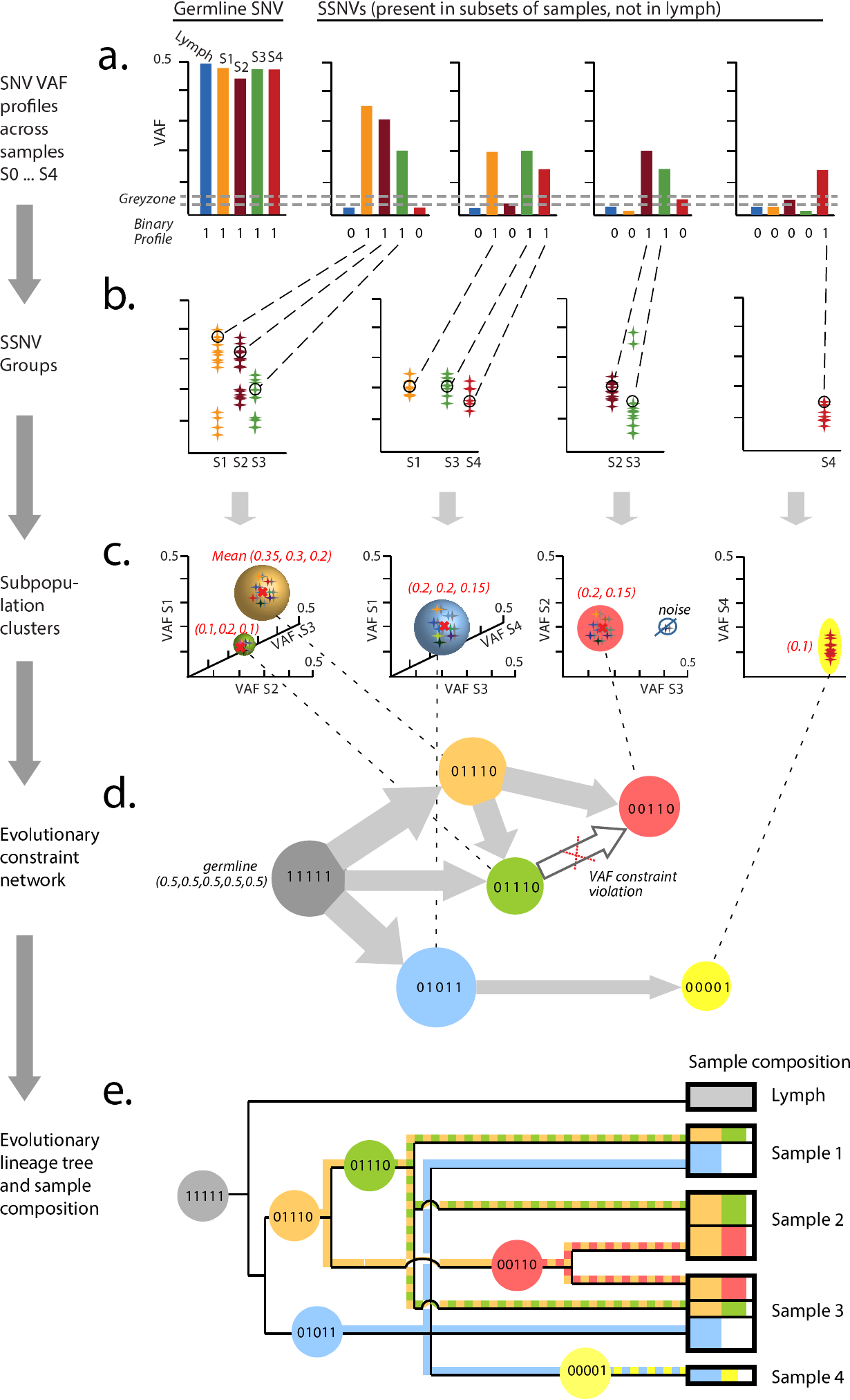}
\caption{(a) Toy example of five samples (Lymph control and four tumor samples S1-- S4) with one germline SNV and four SSNVs, each associated with a binary sample profile. (b) SSNVs are partitioned into groups based on their binary profile (displayed groups contain other SSNVs with varying VAFs). (c) Group SSNV are clustered based on their VAFs. (d) The clusters of each SSNV group are incorporated into an evolutionary constraint network. An edge is placed between a parent and child node if, for each sample, the VAF of the parent is greater than or equal that of the child node. For example, this constraint is violated for the green 01110 node and node 00110. (e) The lineage tree is constructed from the constraint network. The leaves of the tree are the individual samples. The horizontal line subdivisions in a sample indicate mixed lineages, separating the different subpopulations of cells in the sample. The colors in each subdivision describe the mutation groups that the cells in this subpopulation have. Consider the orange node 01110, which denotes SSNVs of class 01110. Those SSNVs are found in samples 1, 2, and 3. After an ancestral cell division, the daughter cells' lineages accumulated SSNVs too (01110, green; 00110) that are now present in their descendant samples or subclones. About 20\% of sample 1 are cells that come from the orange and green lineage, and about 40\% come from the blue lineage. Samples 1, 2, and 3 grew from two or more subclones, whereas sample 4 only grew from one subclone.}
\end{figure*}

At a high level, the LICHeE algorithm can be broken down into the following main steps (Figure 1). First, LICHeE partitions SSNVs into groups based on their occurrence in each sample, such that each group stores all the mutations that were called in the same subset of samples. To separate subclone lineages, the SSNV members of each resulting group are then further clustered based on their VAFs, such that SSNVs with similar VAFs across samples are clustered together. The final lineage tree needs to provide a valid ordering of these resulting SSNV clusters (i.e. an ordering that does not violate the three constrains defined above). In order to find such a tree, we construct an \textit{evolutionary constraint network}, which encodes whether a given cluster of SSNVs could have preceded another, for each pair of clusters. More specifically, this network is an acyclic directed graph (DAG) that has SSNV clusters as its nodes and whose edges encode possible predecessor relationships among the nodes' mutations (i.e. an edge denotes that the mutations of a given pair of clusters satisfy ordering constraints (1) and (2)). This network greatly reduces the search space of possible valid trees and allows us to formulate the task of inferring such trees as a search for spanning trees of the network that satisfy constraint (3) (within a given error margin), which ensures that the reconstructed trees are composed of parent-daughter edges that exhibit somatic VAF consistency with the cell lineage expansion. If multiple valid lineage trees are found during the search, the trees are ranked based on how well they support the cluster VAF data, the top-ranking tree minimizing the use of the permitted error margin (see Methods for details). Finally, as shown in Figure 1, the leaves of the resulting trees are the individual samples (added post-search), whose composition can be reconstructed by tracing back their respective subclone cell lineages in the tree. We detail each of these steps in Methods.

Since finding true SSNV groups is a crucial step of the algorithm, it is important to minimize false positive and false negative SSNV calls across samples. However, accurately detecting SSNVs in each sample is a challenging task due to high levels of noise in the data, which can come from various sources, such as sequencing errors, systematic amplification bias, mapping errors, and sample impurities. Multiple techniques have been developed to address this problem to date  \cite{gerstung2012reliable,yost2012mutascope,larson2012somaticsniper,koboldt2012varscan,saunders2012strelka}, many employing a Bayesian approach to model the distributions of noise and true genotypes in matched-normal samples. LICHeE can work with variant calls produced for each sample by any specialized existing method. However, it does not require the users to pre-process the data using these tools, providing its own heuristic mechanism to call SSNVs using the multi-sample VAF data. At a high level, it first finds SSNVs that can be called reliably in each sample using two hard thresholds $T_{present}$ and $T_{absent}$, above and below which, respectively, the SSNV are considered robustly present or absent. Then, assuming that the presence patterns of such SSNVs capture most of the topology of the true underlying evolutionary tree, it uses this inferred tree to inform the group assignment of the SSNVs whose VAF falls in between the thresholds (the 'greyzone').

Currently LICHeE does not automatically detect or incorporate the CNVs explicitly into the model, although the method can still find valid phylogenies even in the presence of SSNVs within such regions (for example, each patient in the ccRCC dataset had numerous variants from CNV regions). In order to address this limitation, LICHeE also accepts cell prevalence (CP) values instead of VAFs, which can be computed by several recently-developed tools, such as PyClone \cite{roth2014pyclone}, ABSOLUTE \cite{6}, and ASCAT \cite{van2010allele}, and account for CNVs, LOH status, and sample purity. The same algorithmic steps can then be directly applied to CP values of each variant. It can be easily seen that the three perfect phylogeny ordering constraints still hold for CP values and can be used to search for the underlying lineage tree. Furthermore, in order to support outputs of specialized clustering approaches (e.g. PyClone \cite{roth2014pyclone}), LICHeE also accepts already computed clusters of mutations (with given CP and VAF-based centroid values) and uses these clusters as nodes of the phylogenetic constraint network. 

We evaluated LICHeE on three recently published ultra-deep sequencing multi-sample datasets of clear cell renal cell carcinoma (ccRCC) by Gerlinger et. al  \cite{15}, high-grade serous ovarian cancer (HGSC) by Bashashati et. al  \cite{19}, and breast cancer xenoengraftment by Eirew et. al \cite{eirew2014dynamics}, as well as on simulated trees of heterogeneous cancer cell lineage evolution. On the ccRCC dataset, LICHeE constructs near-identical trees to the trees published in the study. We show that the interesting difference in the topology of one tree arises due to potential heterogeneity of a sample that cannot be discovered using the traditional maximum parsimony approach and analyze when this approach can fail to detect existing sample subclones. For each patient LICHeE finds a unique valid tree. On the HGSC datasets we show that the trees generated by LICHeE are better supported by the data and demonstrate why applying neighbor joining with Pearson correlation distance metric, used by the study, might not be suitable for cancer datasets. LICHeE finds a unique valid tree for all patients except Case5. Finally, we show that the trees inferred by LICHeE on the xenoengraftment dataset are consistent with the single-cell analysis done by the study. On each patient dataset LICHeE takes only a few seconds to run.

\subsection*{Lineage Tree Reconstruction on ccRCC Data}

\begin{figure*}[t!]
\includegraphics[scale=.5]{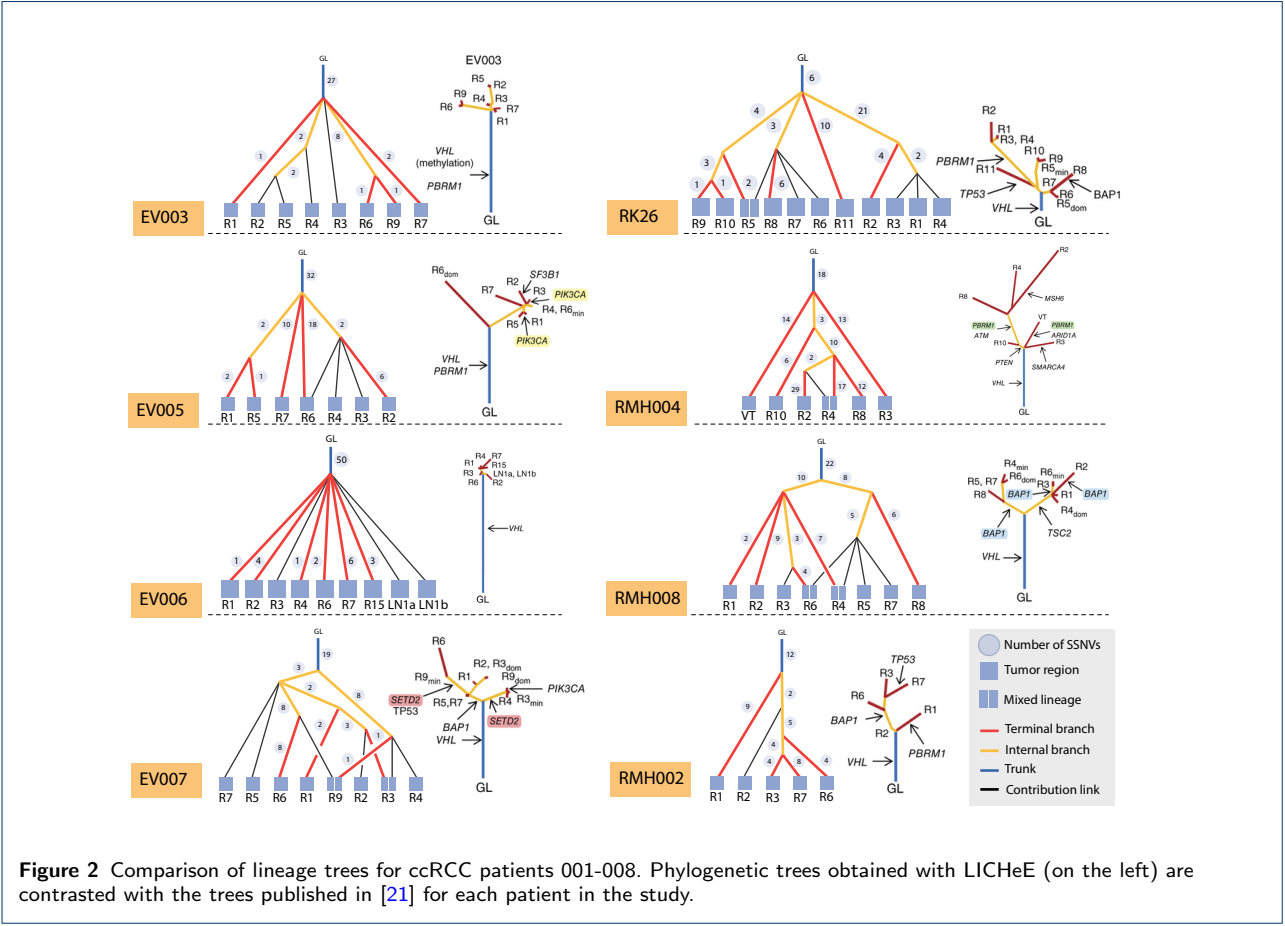}
\caption{Comparison of lineage trees for ccRCC patients 001-008. Phylogenetic trees obtained with LICHeE (on the left) are contrasted with the trees published in \cite{15} for each patient in the study.}
\end{figure*}

The ccRCC study by Gerlinger et. al  \cite{15} validated 602 nonsynonymous nucleotide substitutions and indels from multiple samples of 8 individuals. It used VAF-based clustering of each sample to detect subclones prior to determining the variant presence patterns used in tree reconstruction. The phylogenetic trees were then reconstructed using maximum parsimony, revealing a branched pattern of ccRCC evolution in all tumors. Figure 2 juxtaposes the trees produced by LICHeE with the trees presented by Gerlinger et. al  \cite{15} (for details about the parameters used by LICHeE and the ccRCC dataset see Appendix B). The trees presented in the figure have been redrawn using AI.  We can see that the trees generated by LICHeE are topologically \emph{identical} (consisting of the same branches) to the published trees for patients EV005, EV007, RMH002, RMH008, and RK26. Furthermore, LICHeE identified subclones in all the samples reported to be heterogeneous by the study. In particular, it identified the following regions as a mixture of two subclones: R6 in EV005 (with frequencies of 0.29 and 0.04), R3 and R9 in EV007 (with frequencies of 0.15 and 0.03 and 0.21 and 0.03, respectively), R4 and R6 in RMH008 (with frequencies of 0.19 and 0.14 and 0.21 and 0.15, respectively), and R5 in RK26 (with frequencies of 0.15 and 0.03). These subclones correspond to the dominant (dom) and minor (min) shown in the published trees. Evidence supporting each subclone can be analyzed using the presented trees. 

The trees generated for patients EV003 and EV006 also highly match the published results. For EV006, the LICHeE-generated tree does not contain the following two partially shared groups: (R2, LN1a, and LN1b) and (R3, R1, R4, R7, R15). For the first group we find no evidence in the data -- no SSNVs are shared in these three samples and absent from the others. We do find one mutation that supports the second group. Because, by default, LICHeE eliminates nodes that have evidence from only one SSNV, this group is not shown in our tree.

The RMH004 dataset contains three partially overlapping groups (R3, VT, R10, R4, R2), (R3, VT, R10, R4, R8), and (R10, R4, R2, R8). These groups represent separate branches where their lowest common ancestor is the group with mutations present across all samples (R3, VT, R10, R4, R2, R8). However, the VAF of this parent group in sample R10 is 0.34, while the average VAF across each of the three groups is 0.32, 0.27, and 0.21, respectively. Therefore, no more than one of these groups can be a descendant of the parent group, without violating the VAF phylogenetic constraint. In order to generate a valid tree, the two least populated conflicting branches among the three are removed from the dataset. Similarly, these two groups are ignored by the maximum parsimony algorithm and are not present in the published tree. The difference between our tree and the published one comes from the mixed lineage we observe in sample R4. Due to the VAF of 0.17 of the group of R4 private mutations, the private group is not a descendant of the group shared by R4 and R2 since the average VAF of that node is 0.06, which is too small to be a parent to 0.17. Therefore, our method suggests two different subclones in R4. Since the group (R2, R4) is small (total of three mutations) and has a low frequency in R4, the evidence of the two subclones cannot be considered very strong; however, it does constitute a signal in the data for the R4 mixed lineage possibility.

While the ccRCC study does perform VAF clustering of each sample to find subclones at the onset, it runs the phylogenetic reconstruction using the presence pattern profiles only. On the other hand, our tree reconstruction with LICHeE applies the VAF constraint to the resulting tree topologies and clusters SSNVs in each group based on their VAFs across all the samples. For patient RMH004, applying the VAF constraint, reveals additional sample heterogeneity, for which we produced a potentially improved tree compared to the tree reported using maximum parsimony. It can also be shown that clustering across all the samples rather than each sample individually, is a more appropriate approach for revealing the heterogeneity in the data. For example, if two subclones occur with high and low VAFs in one sample but are uniform in another sample, single sample clustering will detect them only in the sample where they differ. For example, if the two subclones in samples R3 and R9 of patient EV007 (discussed above) had highly similar VAFs in these samples, clustering would not be able to differentiate the subclones. On the other hand, LICHeE would still be able to detect the mixed lineages in the two samples due to the presence of groups (R1, R2, R3, R5, R6, R7, R9) and (R3, R4, R9) in the data, which must form divergent branches in the lineage tree.

\subsection*{Lineage Tree Reconstruction on HGSC Data}

\begin{figure*}[t!]
\includegraphics[scale=.5]{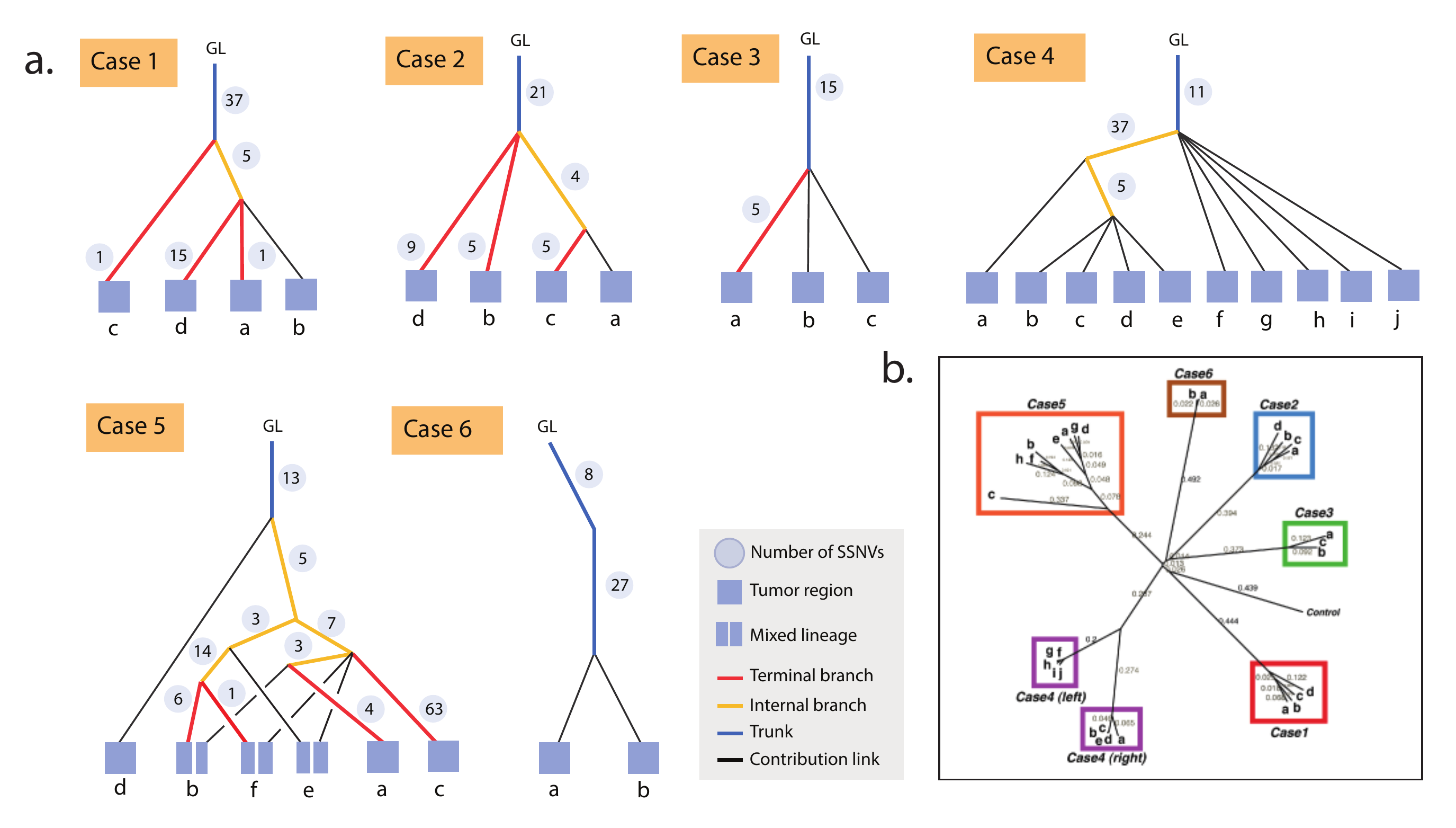}
\caption{Comparison of lineage trees for HGSC cases 1-6. Phylogenetic trees obtained with LICHeE (Panel a) are contrasted with the trees published in \cite{19} (Panel b) for each patient in the study.}
\end{figure*}

We further evaluated LICHeE on the HGSC dataset from the study by Bashashati et. al.  \cite{19}. This study validated 340  somatic mutations from 19 tumor samples of six patients. It used neighbor joining with Pearson correlation distances (computed on the binary sample presence patterns) to infer lineage trees. The trees generated by LICHeE juxtaposed with those presented in the paper are shown in Figure 3 (for details about the parameters used by LICHeE and the HGSC dataset see Appendix B). In four out of six cases (Case2, Case3, Case 4, and Case6) the possible tree topologies are very simple and the trees produced by the two methods are unsurprisingly highly similar. Below we discuss the difference between the two remaining trees.

For Case1 the tree reported in the paper suggests that sample d diverged first, followed by c. This suggests that there should be a group of mutations shared exclusively between samples a and b and another group shared between samples a, b, and c. However, examining the dataset VAF values, as well as the results of the binomial test determining SSNV presence in the samples used by this study, we found no evidence for these two groups. On the other hand, the tree produced by LICHeE suggests the presence of mutations shared by samples a, b, and d only, which we found both in the results of the study's binomial test and by applying the LICHeE hard threshold caller.

For Case5, the study reports an early divergence of sample c. This suggests that there should be mutations shared between all the samples except c. We have confirmed that no such mutational profile exists in the data. On the other hand, the data shows the presence of mutations that exist in samples a, b, c, e, and f but not in d that cannot be supported by the reported tree. The reason why the neighbor-joining algorithm of the study chose sample c as the first diverging branch of the tree must be because of the large presence of private mutations in sample c, which led to a low Pearson correlation (and hence a greater distance) with the profiles of other samples. This shows that using the Pearson correlation metric is not suitable for this data. Furthermore, directly applying traditional phylogeny reconstruction techniques (e.g. neighbor joining) cannot reveal sample heterogeneity.

The tree produced by LICHeE for Case5, reveals mixed lineages in samples b, e, and f. Interestingly, LICHeE detects two clusters in group (a, b, c, e, f) with mean VAFs of [0.29, 0.34 0.3, 0.19, 0.4] and [0.17, 0.14, 0.13, 0.1, 0.13], respectively. The two subclones in each sample are then produced by the divergence of the higher VAF cluster into two branches: one branch containing the lower VAF cluster and the other branch containing the group of mutations (e, f, b). While the presence of group (e, f, b) is weak (supported by three mutations only), the presence of group (b, f) is substantial (supported by 14 mutations) and provides strong evidence for the mixed lineage of b and f. Furthermore, due to the presence of group (a, b), the group (b, f) cannot be assigned as a child of the low VAF cluster (a, b, c, e, f) without violating the VAF phylogenetic constraint. However, since there are only three mutations in group (a, b), eliminating this group may be a reasonable alternative reconstruction. Therefore, multiple topologies are possible in this scenario, each supported by varying degrees of evidence. LICHeE presents the tree that best fits the data given input parameters of expected noise level and minimum mutational support needed for a node. Using LICHeE with different parameters and interacting with the trees can allow users to explore other evidence existing in the data. It is important to point out that, as opposed to the results of the neighbor-joining algorithm that produces trees with lack of evidential support in some branches, each branch in a tree reported by LICHeE reflects the presence of the corresponding SSNV groups in the data. 

\subsection*{Lineage Tree Reconstruction on Xenoengraftment Data}

\begin{figure*}[t!]
\includegraphics[scale=.35]{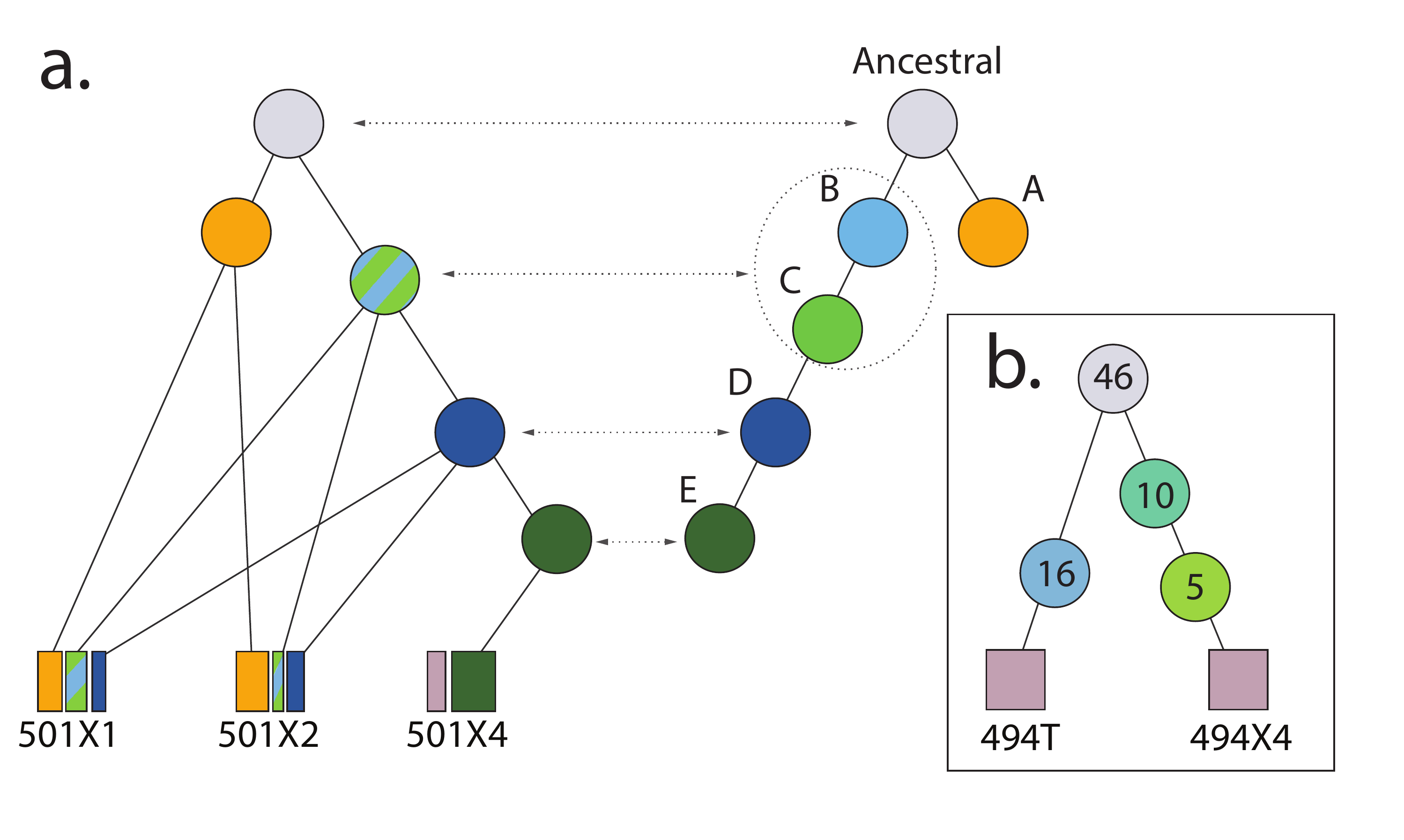}
\caption{LICHeE reconstruction on breast cancer xenoengraftment data \cite{eirew2014dynamics}. a, Lineage tree generated by LICHeE (on left) is juxtaposed with the phylogeny derived from single-cell data in the study (on right) for SA501. b,  Lineage tree generated by LICHeE for SA494.}
\end{figure*}

The study by Eirew et. al. \cite{eirew2014dynamics} used deep-genome and single-cell sequencing to evaluate the clonal dynamics of xenoengraftment of breast cancer tissue into immunodeficient mice. The study applied PyClone \cite{roth2014pyclone} to infer mutational clusters representing clonal genotypes and then validated these results using single-cell analysis of two cases, SA494 and SA501. The study used Bayesian phylogenetic inference to reconstruct the evolutionary relationships between the single-cell nuclei. We applied LICHeE to the deep-sequencing VAF data of this study. We then compared the results of LICHeE to the single-cell phylogenetic trees and clonal genotypes for both SA494 and SA501. 

Single-cell phylogenetic inference of SA501 passages X1, X2,  and X4 reveals an ancestral genotype that branched into two sibling clades A and B. Sequential acquisition of additional mutations in clade B then gave rise to genotypes C, D, and E. Samples X1 and X2 were found to be a mixture of clones with genotypes A, B, C, and D; while, sample X4 was found to be dominated by genotype E and did not contain clones with genotype A (see Figure 2 in \cite{eirew2014dynamics}). As can be seen in Figure 4a, the tree reconstructed by LICHeE mirrors the phylogeny and sample compositions revealed by the study. In particular, it presents the same two sibling clades derived from the ancestral genotype. Samples X1 and X2 both contain subclones with genotype A (with a higher percentage of this genotype in sample X2, as confirmed by the single-cell analysis), while this genotype is missing from sample X4. Similarly, genotype E is found to be private to and dominates sample X4. The only difference in the LICHeE tree is the collapse of genotypes B and C into one cluster; however, examining the bulk CP values reported by the study, we can see that the CP values of these genotypes are highly similar in the three given samples (X1, X2, and X4) and, as a result, cannot get subdivided into two separate clusters during the clustering step (note: the PyClone analysis of the study was simultaneously performed on three additional samples T, X3, and X5, which showed higher CP value differences).   

The single-cell analysis of SA494 samples T and X4 reveals two clades, with mutually exclusive sets of mutations, emerging from an ancestral clone present in both samples (see Extended Data Figure 3 in \cite{eirew2014dynamics}). LICHeE reconstructs the exact same topology (Figure 4b), showing two groups of mutations private to T and X4, respectively. In accordance with the results reported by PyClone, LICHeE also finds two clusters of mutations private to X4 (with the descendant cluster present in about 20\% of the sample). 

\subsection*{Simulations}

We developed a cancer cell lineage simulator to better asses the performance of LICHeE. Our simulator models cancer evolution from normal tissue producing a branching hierarchy of monoclonal cell populations in accordance with the branched-tree cancer evolution model \cite{greaves2012clonal, gerlinger2012intratumor, yap2012intratumor}. Starting with the normal cell population, the simulator iteratively expands the cell lineage tree by introducing (with some given probability) new daughter cell populations corresponding to newly acquired SSNV or CNV events. In particular, in every iteration, each cell population present in the tree can give rise to a new population of cells (with a given randomly generated size) representing a new SSNV with probability $P_{SSNV}$ or a CNV event with probability $P_{CNV}$. Each cell population can also undergo a cell death event with probability $P_{Death}$. Each simulated SSNV is randomly associated with a genome location (chromosome and position) and haplotype; the CNVs are associated with a chromosome arm and haplotype and correspond to a duplication of this chromosome arm. For the evaluation, we generated 100 lineage trees, each expanded over 50 iterations, with the following parameters: $P_{SSNV} = 0.15$; $P_{CNV} = 0$, $0.1$, and $0.18$; and $P_{Death} = 0.06$. This process results in lineage trees with an arbitrary number of branches and nodes (several hundred to thousands of nodes on average). Figure 5 illustrates one such lineage tree. 

Multiple samples are then collected from each lineage tree. Each sample consists of several cell populations (nodes) of the tree, where each such cell population represents a subclone in the sample. We implemented two sampling schemes: $randomized$ sampling and $localized$ sampling. The randomized sampling process selects a random subset of nodes from the tree for each sample; on the other hand, the localized sampling process is meant to mimic biopsies from spatially distinct sites and selects nodes such that samples mostly contain subclones from distinct branches of the simulated tree. Localized sampling for $n$ samples is achieved by selecting $n$ disjoint subtrees in the simulated tree using an approach based on breadth-first search, which finds disjoint subtree roots as high on the tree as possible (if $n$ disjoint subtrees do not exist in the tree, the maximum number of subtrees is used and some samples are obtained from the same subtree in a round-robin fashion). Figure 5 illustrates the localized sampling procedure for 10 samples.

\begin{figure*}[t!]
\includegraphics[scale=.06]{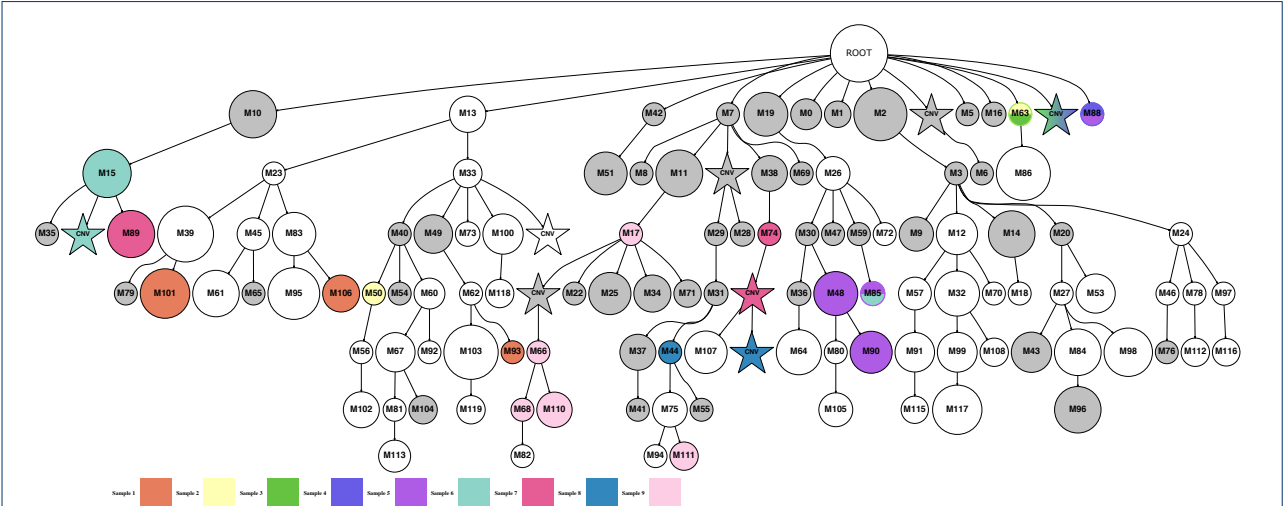}
\caption{Example of a small simulated lineage tree (with SSNV and CNV events) with localized sampling (10 samples total indicated by different colors). Each node represents a cell population that contains every mutation in its lineage. Gray indicates dead cell populations.}
\end{figure*}

Given a selected subset of cell populations, the sample is then created by obtaining a fraction of the cells from each population by sampling from a multinomial distribution with probabilities corresponding to the cell population sizes. For randomized sampling, we select up to 5 subclones for each sample. For localized sampling, there can be up to 5 subclones from the same distinct subtree, exactly one subclone from a neighboring subtree, and some fraction of normal cells to represent normal contamination (the fraction is randomly selected to be from 0 up to 20\% of the sample).  In order to determine how the performance of LICHeE degrades as the number of SSNVs located in CNV regions grows, we ran several experiments with increasing numbers of CNV events. In particular, we set $P_{CNV}$ = [0, 0.1, 0.18], which resulted in a total of 0\%, 65\%, and 80\% of SSNVs to be affected by CNVs in the collected samples, respectively. Given the sample cell population counts, we can compute the VAF of each SSNV in this sample. In the absence of CNVs, the VAF of a given SSNV is simply the fraction of the cells containing the SSNV out of all the cells in the sample. When CNVs are present, we count the number of haplotypes containing the SSNV and the reference allele across all cells of the sample. Given an SSNV, we consider the chain of mutations present in each cell population affecting its genome position. Let $H_P^v$ be the number of haplotypes containing the SSNV in population $P$ and $H_P^r$ be the number of haplotypes containing the reference allele. The VAF of a given SSNV $M$ is then:
\begin{center}
$VAF(M)=\frac{\sum_{P=1}^N {P_n \cdot H_P^v }}{\sum_{P=1}^N{P_n \cdot (H_P^v + H_P^r)}} $,
\end{center}
where $N$ is the total number of selected populations and $P_n$ is the number of cells selected from population $P$.

Finally, given the true VAFs of each SSNV, we add sampling and sequencing noise to each value. In particular, in order to simulate reads covering each SSNV position, we sample the VAFs from the Binomial distribution $B(n,p)$, where $p$ is the true VAF of each SSNV and $n$ is the total simulated read coverage (100X, 1,000X, and 10,000X); the generated frequencies have a mean $p$ and variance $\frac{p(1-p)}{n}$. We simulate a Q30 (1 in 1000) base call sequencing accuracy. 

Given the simulated VAFs, we first assessed the performance of LICHeE in SSNV calling. Table 1 presents LICHeE's sensitivity in calling mutations across the samples (i.e. the number of SSNVs with correctly identified sample presence patterns out of all the simulated SSNVs) given true VAF values and for coverages of 100x, 1,000X, and 10,000X. As expected, higher coverage results in higher sensitivity. The method achieves $94-99\%$ sensitivity across all the experiments. 

Next we compared the topology of the reported lineage trees and the simulated trees and measured LICHeE's accuracy in reconstructing the ancestor-descendant and sibling relationships. For every pair of mutations in the simulated tumor hierarchy, we checked if LICHeE preserved the relationship between them in the generated top tree. More specifically, we checked for the following two types of violations: 1) if an ancestor-descendant relationship was inverted or became a sibling relationship, and 2) if a sibling relationship became an ancestor-descendant relationship. Tables 2 and 3 summarize the accuracy of these metrics. Since LICHeE may remove nodes from the network representing non-robust SSNV groups (as described in Methods) if no valid trees are found during the search, not all of the simulated mutations will be present in the final tree. We report the percentage of the SSNVs and of simulated ancestor-descendant (AD) and sibling (Sib) mutation pairs that are present in the tree. As expected, the number of mutations present in the trees decreases with higher numbers of input samples, lower coverage, and the presence of CNVs (which can significantly alter the VAF values, causing the violation of the VAF ordering constraints). In particular, with a low coverage of 100X and 15 samples, only 81\% of SSNVs are preserved in the tree. Therefore, LICHeE is best applied to data with higher coverage when the number of input samples is high. For instance, with a higher coverage of 1,000X, 91-94\% of SSNVs and 83-88\% of mutation pairs are present in the trees with 15 samples (although this metric drops with the presence of CNVs to 91-92\% SSNVs and 66-70\% SSNV pairs).

Since LICHeE groups mutations with the same presence patterns across samples and similar VAFs, only the mutations occurring in a different set of samples or with significantly different VAFs will be placed in distinct nodes of the tree. We report the percentage of such ancestor-descendant pairs (AD-Ord) in Tables 2 and 3. We then evaluate how many ordered mutations preserved the correct ancestor-descendant relationship (AD-Corr). Across all the experiments without CNVs, we get 99-100\% correctness (with less than 1\% of pairs being in reversed order). We see 92-96\% correctness in experiments with 80\% of SSNVs located in CNV regions and 1000X coverage. AD mutations that were not ordered or grouped in the same node, will be siblings in the reconstructed tree (AD $\rightarrow$Sib). We find that the vast majority of such mutation pairs involve private mutations, whose placement in the tree is usually under-constrained (i.e. multiple tree nodes can serve as ancestors to such mutation groups). Finally, we can see that the reverse violation of sibling mutations being placed into ancestor-descendant nodes (Sib$\rightarrow$AD) is also very rare (up to 7\% across all the experiments). Therefore, we conclude that the trees reconstructed by LICHeE provide highly accurate ordering of the mutations in its nodes. 

\section*{Conclusion}

LICHeE has been designed to automatically infer cell lineages of multiple tumor samples and the sample subclone decomposition. Our analysis shows that LICHeE is highly effective in reconstructing the phylogenies and uncovering the heterogeneity of previously published datasets and in simulations, improving not only upon traditional tree-building methods, but also on recent developments specialized for cancer data. Currently LICHeE works with deep sequencing data that provides VAF estimates with low variance (as well as on CP values that can be obtained from existing tools and can correct for SCNAs, LOH, and sample purity). SSNV data obtained from deep whole-genome sequencing, targeted resequencing of informative SSNVs, or exome sequencing in tumors with a high degree of somatic SSNVs present in exomes, should be appropriate inputs to LICHeE. Several directions for future work are open: extension of this method to lower coverage whole-genome sequencing data and incorporation of aneuploidies and large CNVs directly into the model.

\section*{Methods}

\subsection*{Grouping and Clustering SSNVs}
When partitioning SSNVs into groups based on sample occurrence, each SSNV is first associated with a binary sample profile denoting its presence or absence in each sample. Given $S$ samples from an individual, the \textit{binary profile} is defined as a binary sequence of length $S$ where the $i^{th}$ bit is set to 1 if this SSNV is called in the $i^{th}$ sample, and is 0 otherwise (e.g. given 5 samples, an SSNV with the profile 01011 is called in samples 2, 4, and 5). SSNVs with the same profile are assigned to the same SSNV group (e.g. a group with the profile 01101 will contain all the SSNVs occurring in samples 2, 3, and 5). The group with the profile consisting entirely of 1s will contain SSNVs that occur in all the samples and are, therefore, germline variants (assuming that the sample set includes a normal control sample).

\begin{figure*}[t!]
\includegraphics[scale=.25]{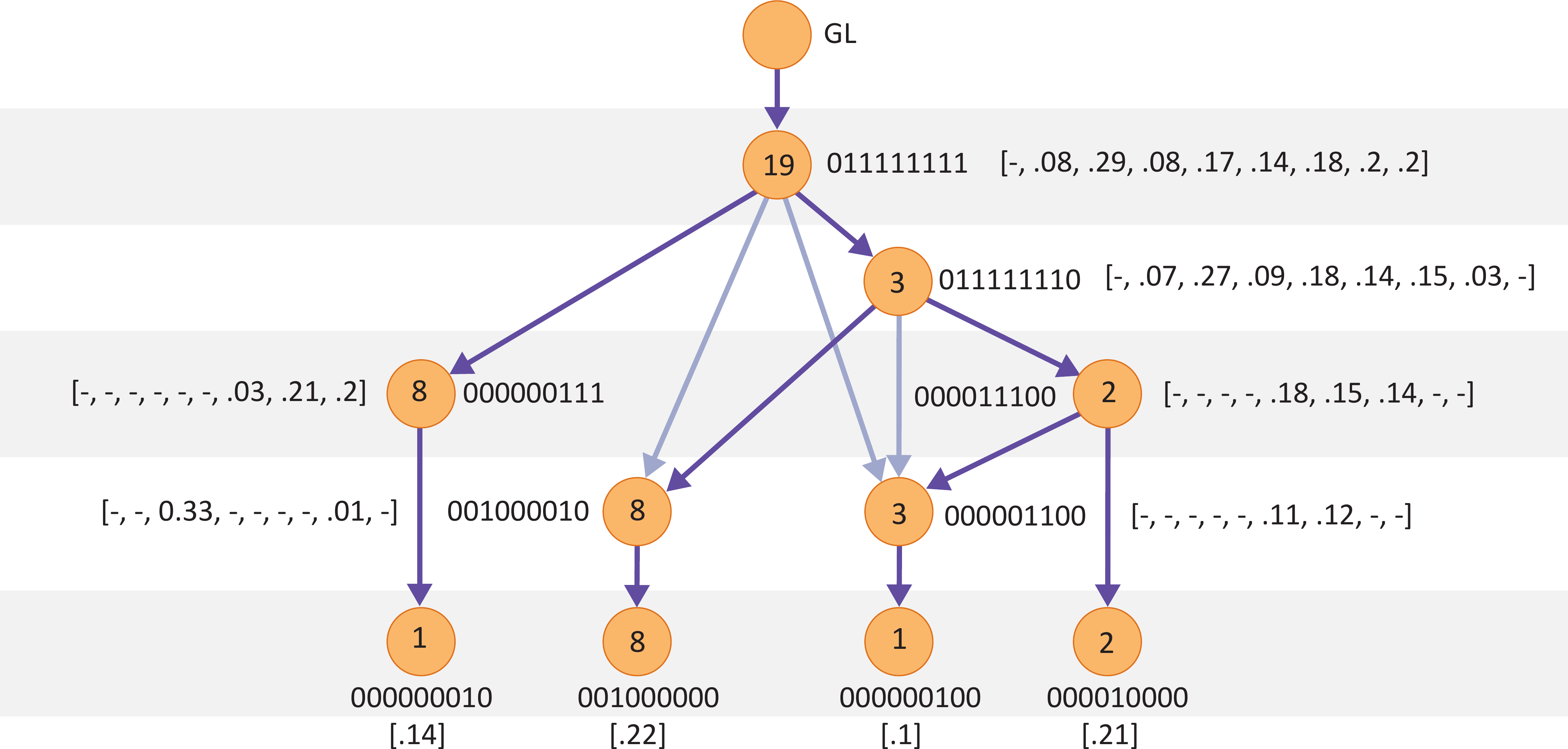}
\caption{Evolutionary constraint network for patient EV007 with 8 tumor samples and 55 SSNVs. Each node is associated with the binary profile of its corresponding SSNV group, the VAF centroid, and the number of SSNVs assigned to it. The edges represent the potential precedence relationships between the node SSNVs. The spanning tree reported for patient EV007 is highlighted (see Results).}
\end{figure*}  

The SSNV binary profiles can be passed as input or computed from SSNV VAFs as follows. First two hard VAF thresholds, $T_{present}$ and $T_{absent}$ are used to determine if an SSNV is robustly present or absent from a sample. An SSNV profile is classified as robust if its VAF is above or below the two thresholds, respectively, and if at least a minimum number of other robust SSNVs (default set to one) have the same binary profile. All other SSNVs are considered non-robust and are assigned to a group as follows. Given a non-robust SSNV $m$, its VAF across samples can either fall below (marked 0), above (marked 1), or in between (marked *) the thresholds $T_{present}$ and $T_{absent}$, resulting in a profile such as 01*11. The candidate groups, to which $m$ can be assigned, must have an identical profile in all the samples that are marked 0 or 1 (e.g. for profile 01*11 two valid candidate groups are 01111 and 01011). Since $m$ can be assigned to more than one target group, we consider that the group containing a robust SSNV that is most similar in VAF to $m$ is the best candidate. The following metric is used to compute the similarity between two SSNVs $m$ and $n$:
\begin{equation}
{\mathrm{sim}_{mn} = \sum_{i \in \mathrm{samples}}{\frac{\min(m.\mathrm{VAF}_i, n.\mathrm{VAF}_i)}{\max(m.\mathrm{VAF}_i, n.\mathrm{VAF}_i)} \enspace}} \enspace ,
\end{equation}
\noindent where $m.\mathrm{VAF}_i$ is the VAF of $m$ in sample $i$. If the maximum similarity is higher than a given threshold, $m$ is assigned to the group of $\mathrm{argmax}_{n \in \mathrm{Candidates}}{\mathrm{sim}_{mn}}$. Unassigned non-robust SSNVs will form new profile groups. We minimize the number of such new groups by formulating this task as a set cover problem. In particular, let $X$ be the set of all unassigned SNVs. Denote $Y$ as the set of subsets of $X$, where each subset represents mutations that can be assigned to the same potential target group. The target groups are a list of all possible binary profiles that the SSNVs can be assigned to, obtained by substituting all *s by 0 or 1.  We want to find the minimum number of target groups (i.e., smallest subset of $Y$), which cover all mutations in $X$. This problem is known to be NP-complete and searching for the exact solution is not feasible. Instead we apply the standard greedy algorithm by choosing (at each stage) the target group that covers the largest number of non-robust SSNVs. For SSNVs whose targets are not supported by any other mutations, we convert each * to 1 or 0 depending on whether the VAF is closest to $T_{present}$ or $T_{absent}$, respectively. 

Once the SSNVs are partitioned into groups, the SSNVs of each group are further clustered based on their sample VAFs. Each SSNV group is associated with a matrix $M$ of VAFs of size $n \times s$, where $n$ is the number of SSNVs in this group and $s$ is the number of represented samples (e.g. $s = 3$ for a group with the profile 0110001). The EM clustering algorithm for GMMs is run on the resulting VAF matrix $M$ using  \cite{hall2009weka}. The result is a set of SSNV clusters with an associated VAF centroid vector, $\vec{\overline{VAF}}$. To handle the high variance in the VAF data due to noise, some of the resulting clusters are eliminated (based on size) or collapsed with neighboring clusters, based on the distance between their centroid vectors.

\subsection*{Evolutionary Constraint Network Construction}

Given the clusters of each SSNV group, we construct an evolutionary constraint network to capture valid evolutionary timing relationships between the mutations of each cluster pair. The network is a DAG, where each node corresponds to an SSNV cluster (except the root, which represents the germline) and each edge between two nodes, ($u \rightarrow v$) denotes that parent node $u$ could be an evolutionary predecessor of child node $v$ (i.e. that SSNVs in cluster $u$ could have "happened before" SSNVs in cluster $v$). In particular, an edge ($u \rightarrow v$) is added only if the nodes satisfy the following two constraints $\forall i \in \mathrm{samples}$ (which guarantee that the network will be acyclic): 
\begin{equation}
\begin{array}{l}
(1)\enspace u.\mathrm{\overline{VAF}}_i \geq v.\mathrm{\overline{VAF}}_i - \epsilon_{uv} \mbox{ and } \\
(2)\enspace \mbox{if } u.\mathrm{\overline{VAF}}_i = 0, v.\mathrm{\overline{VAF}}_i = 0 \enspace ,\\
\end{array}
\end{equation}

\noindent where $\epsilon_{uv}$  is the VAF noise error margin (note: the error margin is the maximum of the sum of the standard errors for sample $i$ of the two clusters and a configurable parameter). In the resulting network, each node will have at least one parent (since all nodes can be connected to the root). We avoid checking all node pairs, by organizing the nodes into levels according to the Hamming weight (i.e. number of 1s) of the binary group profile to which they belong. Nodes that are in the same level and have conflicting binary profiles cannot satisfy the above constraints. Nodes from different levels can only be connected such that the node in the higher level is the parent of the node in the lower level. Finally, for nodes that are in the same level and have the same binary profile, the edge is added in the direction that minimizes $\mathrm{VAF}_{\mathrm{ERR}}$, where:
\begin{equation}
\mathrm{VAF}_{\mathrm{ERR}_{u \rightarrow v}} = \sum_{i \in samples}{\vec{1}_\mathrm{ERR} \cdot (v.\mathrm{\overline{VAF}}_i  - u.\mathrm{\overline{VAF}}_i)^2}
\end{equation}
with the indicator function:
\begin{equation}
\vec{1}_\mathrm{ERR} = \left\{
\begin{array}{l}
1, v.\mathrm{\overline{VAF}}_i > u.\mathrm{\overline{VAF}}_i \\
0, \mbox{ otherwise }
\end{array}
\enspace .
\right.
\end{equation}
Figure 6 illustrates the constraint network produced for the dataset of ccRCC Patient EV007 (described in the Results section). 

\subsection*{Phylogenetic Tree Search}

By the constraint network construction, a valid lineage tree $T$ of the SSNV clusters (i.e. a tree that does not violate the three constraints of the perfect phylogeny model) must be a spanning tree of the network that satisfies the following requirement $\forall \mbox{ nodes } u \in T$:
\begin{equation}
\forall i \in \mathrm{samples \enspace :} \sum_{v \mbox{  s.t. }(u \rightarrow v) \in T}{v.\mathrm{\overline{VAF}}_i} \leq u.\mathrm{\overline{VAF}}_i+ \epsilon.
\end{equation}
\noindent That is, the sum of the VAF centroids of all the children must not exceed the centroid of the parent. We use inequality here since our method does not require all the true lineage branches to have been observed. To tolerate noise in the VAF data, we relax the constraint by allowing the sum of children VAFs to exceed the parent by an error margin $\epsilon$. Given such a valid lineage tree, each sample can then be decomposed into subpopulations by enumerating all the paths in the tree starting with the germline root and ending in the last node containing mutations in that sample.

The problem of finding all such trees is equivalent to the problem of finding \textit{all spanning trees} of the constraint network DAG for which Eqn. (5) holds. We have extended the Gabow and Myers spanning tree search algorithm  \cite{gabow1978finding} to generate all such spanning trees. The original algorithm generates all spanning trees of a directed graph using backtracking and an efficient bridge edge detection method based on DFS. Our extension consists of enforcing Eqn. (5) during the tree search by terminating the expansion of a given tree and backtracking as soon as an edge violating Eqn. (5) is added (see Algorithm 1). Same as the original algorithm, this search runs in $O(|V| + |E| + |E|N)$ time, where $|V|$ is the number of nodes, $|E|$ is the number of edges, and $N$ is the number of spanning trees in the network. While the runtime of the program depends on the number of spanning trees in the constraint network, in practice the search is very fast (taking on the order of a few seconds). However, since, in theory, it is possible for the algorithm to take longer on datasets that result in constraint networks with many spanning trees, we provide a bound on the maximum number of lineage trees to generate in order to avoid searches that are too long. Similarly, we also have a high bound on the number of calls to the GROW procedure in Algorithm 1. We expect this scenarios to be very rare in typical validation datasets. To reduce search space we also optionally constrain the placement of private mutation nodes in the constraint network to their closest valid predecessors. 

\renewcommand*\Call[2]{\textproc{#1}(#2)}
\begin{algorithm}[!t]
\caption{Finding All Lineage Trees}\label{euclid}
\begin{algorithmic}[1]
\State {Initialization: $f \gets$ emtpy list, $L \gets$ null // stores the last tree output}
\Procedure{Lineage Tree Search}{$N$}  \textit{// N is a constraint network rooted at r}
\State {Tree  $t \gets$ new empty Tree}
\State {$t$.\Call{addNode}{$r$}}
\State{add all edges ($r \rightarrow v$) $\in N$ to $f$}
\State {\Call{grow}{$t$}}
\EndProcedure
\Procedure{grow}{$t$, $N$}
\If{$t$ contains all the nodes in $N$}   
\State {$L \gets t$} 
\State {\textbf{output} $L$}
\Else 
\State {$s \gets$ emtpy list}
\State {$b \gets$ \textbf{false}}
\While{(\textbf{not} $b$ and  $f$ \textbf{not} empty)}
\State{\textit{// e defined as (e.From $\rightarrow$ e.To)} }
\State {Edge $e \gets f.$\Call{removeLast}{}}  
\State {Node $v \gets e.$To}
\State {$t$.\Call{addNode}{$v$}}
\State {$t$.\Call{addEdge}{$e.$From $\rightarrow$ $v$}}
\State{\textit{// ret. \textbf{true} if Eqn. (5) is satisfied for node e.From} }
\If{$t.$\Call{checkConstraint}{$e.$From} } 
\State {add all edges ($v \rightarrow w$), w $\not\in$ $t$ to $f$} 
\State {remove all edges ($w \rightarrow v$), w $\in$ $t$ from $f$} 
\State {\Call{grow}{$t$}}
\State {\textbf{if} number of returned trees $>$ \textit{max\_trees} \textbf{return}}
\State {remove all edges ($v \rightarrow w$), w $\not\in$ $t$ from $f$} 
\State {add all edges ($w \rightarrow v$), w $\in$ $t$ to $f$} 
\EndIf
\State {$t.$\Call{removeEdge}{$e.$From $\rightarrow$ $e.$To}}
\State {$N.$\Call{removeEdge}{$e.$From $\rightarrow$ $e.$To}}
\State {$s.$\Call{add}{e}}
\If {$\exists$ an edge ($w \rightarrow v$) s.t. $w$ \textbf{not} a descendent of $v$ in $L$}
\State {$b \gets$ \textbf{false}}
\Else { $b \gets$ \textbf{true}}
\EndIf
\EndWhile
\ForAll {edges $e$ starting from the end of $s$}
\State{remove $e$ from $s$, add $e$ to $f$, add $e$ to $N$}
\EndFor
\EndIf
\EndProcedure
\end{algorithmic}
\end{algorithm}

\begin{figure*}[t!]
\begin{center}
\subfigure[SSNV group selected] {
\epsfxsize=3in
\epsfbox{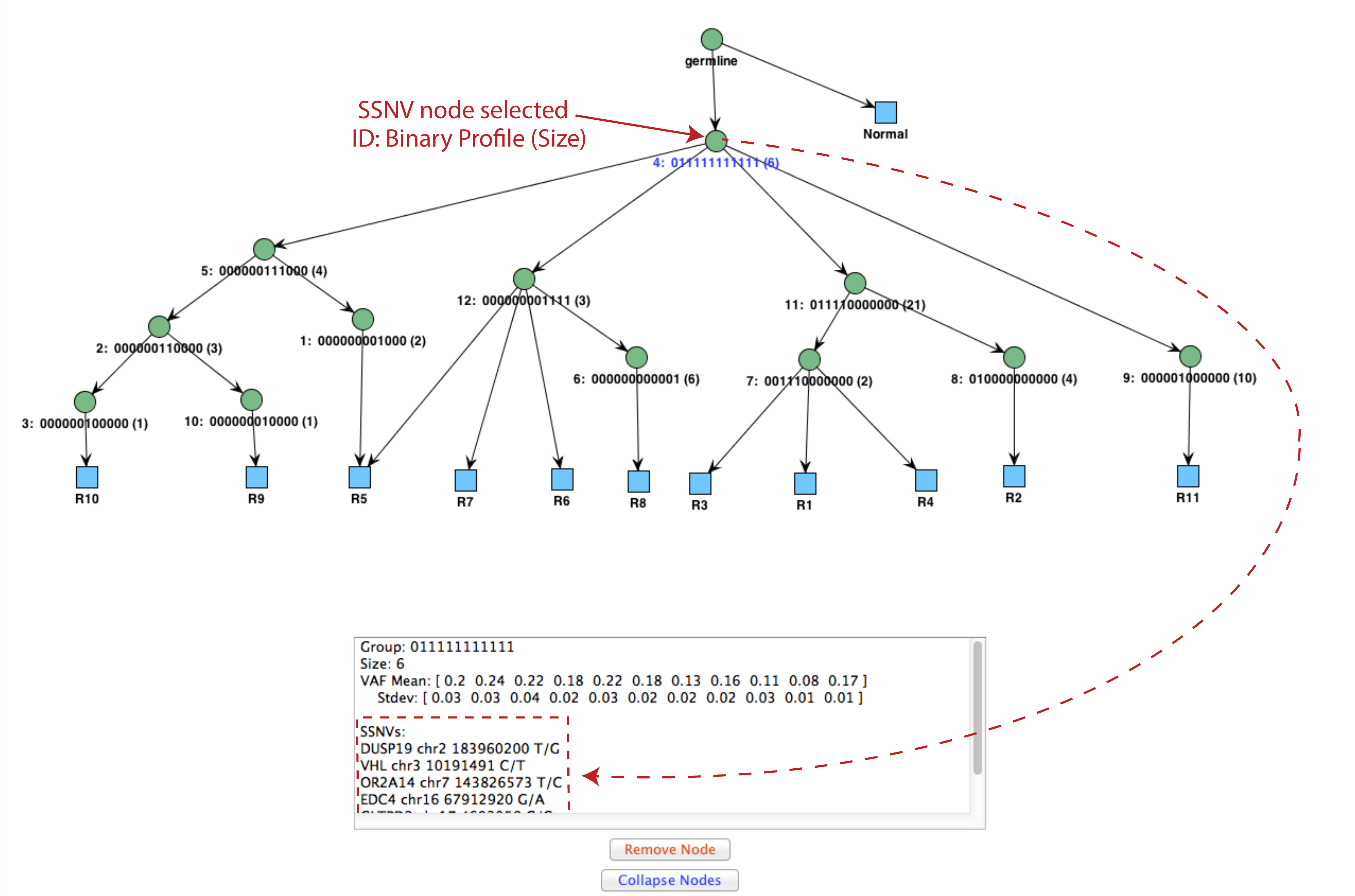}}
\subfigure[Sample selected] {
\epsfxsize=3in
\epsfbox{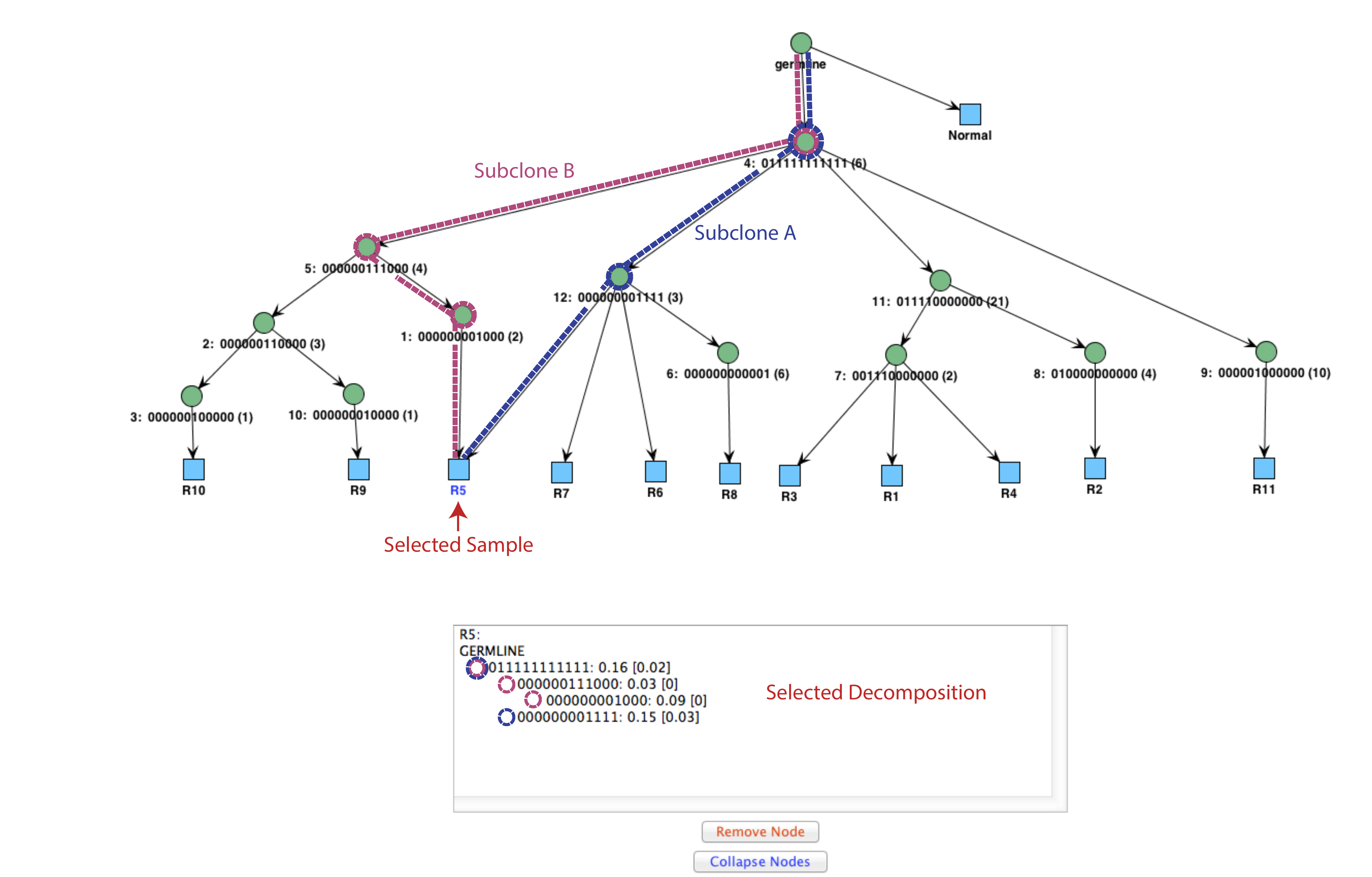}}
\caption{LICHeE graphical user interface output for top lineage tree of ccRCC patient RK26. (a) Information is displayed about the SSNV members of the selected node. (b) Information is displayed about the sample composition.}
\end{center}
\end{figure*}

The above search algorithm will find all spanning trees for which Eqn. (5) holds locally at each individual node; however, it is possible that in order to satisfy this constraint, the centroid values are deviated (within $\epsilon$) in a globally inconsistent way. Therefore, in order to enforce consistency, we apply one additional requirement to the trees returned by the search. In particular, we formulate the following quadratic programming (QP) problem to find a set of $e_{v,i}$ such that for every sample $i$ and node $v$:
\begin{equation}
\begin{array}{l}
\mbox{minimize }  \sum_{v} \sum_{i}  e_{v,i}^2 \\
\mbox{subject to } \\
\sum_{v \mbox{  s.t. } (u \rightarrow v) \in T} \left( {v.\mathrm{\overline{VAF}}_i} + e_{v,i} \right)   \leq u.\overline{\mbox{VAF}_i}  + e_{u,i}   \mbox{  and } \\
 |e_{v,i}| \leq \epsilon, e_{v,i} \leq v.\overline{\mbox{VAF}_i} 
\end{array}
\end{equation}
\noindent If no solution exists, the tree is considered invalid. Since multiple valid lineage trees can be generated, we rank them using the resulting $e_{v,i}$ solution, which corresponds to how well each tree fits the VAF data. The top-ranking tree will be the tree with the minimum sum of squared deviations: $\sum_{v} \sum_{i}  e_{v,i}^2$. Since for certain networks the total number of valid lineage trees can be large, it is impractical to run a QP program on each tree. Therefore, we first rank the trees based on the sum of squared deviations computed locally at each node, and then run QP on the resulting top $k$ trees ($k$ is 5 by default). 

Multiple lineage trees can support a dataset equally well since the placement of certain nodes in the tree may be ambiguous using perfect phylogeny SSNV ordering constraints only (especially when not all lineage branches have been observed). In these cases, more sophisticated custom evaluation criteria would be required to rank the trees. We do not address defining such criteria in this work. Instead we report all the produced trees to the user ranked by their associated score (the number of trees reported is configurable) and provide a GUI to allow the user to easily explore the differences in the topologies (see the Visualization for details). We expect that under any optimality criteria, the sub-optimal trees can also represent signals in the data of potential biological significance. 

It is also possible that no valid lineage trees are found during the search. This can happen if the noise error margin is too narrow or the network contains nodes corresponding to SSNV groups with a misclassified binary profile. In this case the network is adjusted and searched again. Currently the network adjustment procedure will remove one-by-one all the nodes that belong to non-robust SSNV groups (smallest nodes first). Other adjustments, such as increasing the noise error margin, are also possible but not currently implemented.

\subsection*{Visualization}

The constraint network and the phylogenetic trees can be visualized and interacted with in GUI form. The JUNG graph library \cite{o2005analysis} is used to generate the resulting graphs. When visualizing lineage trees, each input sample appears as a leaf in the tree and is connected to the nodes that contain SSNVs present in the sample. By clicking on the nodes of a tree, it is possible to obtain additional information about each node. The information displayed about an internal (non-sample) tree node consists of its binary group profile, its cluster centroid and standard deviation vectors, and the list of the SSNVs in its cluster (these SSNVs can be annotated with information from public databases such as COSMIC, TCGA, etc). If a sample leaf node is clicked, the information displayed consists of the lineage of this specific sample obtained by doing a DFS traversal of the tree starting with the germline root. Finally, the user can rearrange the nodes in the tree, as well as remove nodes and collapse nodes (provided they are clusters of the same group). See Figure 7 for several examples. In addition to the GUI, the program reports the number of trees found and the score of the highest-ranking tree. The user can control how many trees to display.

\subsection*{Implementation}
The LICHeE algorithm was implemented in Java. It is open-sourced and freely available online at \cite{lichee}.

\afterpage{
\clearpage
\begin{landscape}
\begin{table}[t!]
\begin{center}
\label{table1}
\begin{tabular}{|c|c|c|c|c|c|}
\hline
\multicolumn{2}{|c|}{}  & \multicolumn{4}{|c|}{Coverage}  \\  \hline
\# Samples & \# Sim SSNVs & True VAF (+CNV)& 10,000X (+CNV) & 1,000X (+CNV)& 100X (+CNV) \\ \hline
\multirow{2}{*}{5} 
& 26.6 (42.2) &	99.2 (96.9) &98.9 (96.7)	& 99.1 (96.7)	& 97.2 (95.5) \\ 
& 25  (44.5) & 99.5 (97.4)	& 99.1 (97.3)	& 99.3 (97.4)	& 97.4(95.9) \\ \hline
\multirow{2}{*}{10} 
& 43.9 (89.7) &	99.2 (97.8) & 98.5 (97.5)	&98.5 (97.7)	& 95 (96) \\ 
& 40.7 (88.6) & 99.5 (97.5)	& 99.1 (97.3)	& 99	(97.3) & 96.3  (95.9) \\ \hline
\multirow{2}{*}{15} 
& 53.6 (131.1) & 98.8 (96.9)	& 98.2 (96.7) &	97.9 (96.5)	&94.2 (95.4)\\ 
& 51 (136.1) & 99.1(97.6)	& 98.7 (97.4)	& 98.5 (97.2)	& 94.5  (95.8) \\ \hline
\end{tabular}
\caption{SSNV group assignment sensitivity on simulated data. Values indicate the number of correctly assigned SSNVs out of the total number of SSNVs collected in each experiment (\# Sim SSNVs). Results are shown for 5, 10, 15 samples given true VAFs, 10000X, 1000X, and 100X coverage data obtained with localized (top row) and randomized (bottom row) sampling, from trees without CNVs and with ~80\% of SSNVs in CNV regions (in parentheses). All values are averaged over 100 simulated trees. }
\end{center}
\end{table}

\begin{table}[t!]
\begin{center}
\label{table2}
\begin{tabular}{ | l | l | l | l | l | l | l | l | l | l | l | }
\hline
Samples & Cov & Trees & \% SSNVs & \% AD & \% AD-Ord & \% AD-Corr & \% AD$\rightarrow$Sib (-priv) & \% Sib & \% Sib-Corr & \% Sib$\rightarrow$AD (-priv) \\ 
\hline
\multirow{8}{*}{5}
 & \multirow{2}{*}{ t-VAF } 
 & 94	 & 98.9 & 99 & 41.5 & 100 & 30.2 (2.6) & 98.2 & 83.2 & 7.3 (1.5) \\
 & & 98 & 99.5 & 99.5 & 40.4 & 100 & 26.3 & 99.2 & 84.8 & 6.6 (0.9) \\ \cline{2-11}
 & \multirow{2}{*}{10K} 
 & 95 &98.8 & 98.7 & 40.4 & 99.9 & 30.5 (2.5) & 97.8 & 82.5 & 7.7 (1.4) \\
 & & 98 & 99.2 & 98.7 & 39 & 100 & 27.4 & 98.8 & 84.7 & 7 (1.3) \\ \cline{2-11}
 & \multirow{2}{*}{1K} 
 & 95 & 98.6 & 97.7 & 39.7 & 100.0 & 30.5 (2.3) & 97.1 & 82.8	& 7.2 (1.3)\\
 & & 98 & 99.4 & 99.5 & 38.1 & 99.7 & 29.2 & 98.9 & 83.9 & 7.5 (2.2) \\ \cline{2-11}
 & \multirow{2}{*}{100} 
 & 97	 & 96.8 & 93.2 & 38.7 & 99.8 & 31.4 (3.5) & 93.1 & 81.8 & 7.9 (1.7) \\ 
 & & 97 &	97.3 & 94.8 & 35.2 & 99.7 & 30.1 & 95.3 & 83.7 & 7.2 (2.2) \\ 
\hline
\multirow{8}{*}{10} 
& \multirow{2}{*}{ t-VAF } 
&    97 & 97.5 & 96.2 & 58.8 &100 &43.1 (2.2) & 96.9 &94.4 &3.3 (0.2)\\
& & 95 &	97.7	& 96.5&	58.2&	100&	38.1 (1.9) &	96.7&	93.5&	4.2 (1.4) \\ \cline{2-11}
& \multirow{2}{*}{10K} 
& 98 & 96.5 & 94.4 & 57.3 & 100 & 44.9 (4) & 95.2 & 94 & 3.6 (0.4)\\ 
& & 96	& 97 &	96.1 &	56.5&	100&	39.8 (2.3)&	96.2&	93.3&	4.3 (1.2) \\ \cline{2-11}
& \multirow{2}{*}{1K} 
& 98	& 97.4 & 96.4 & 57.5 & 99.9 & 45.2 (3.9) & 96.8 & 94.1 &3.7 (0.4)\\
& & 96 &	97 &	95.7 & 57.4&	99.9&	38.7 (2.4) &	96 &	92.7 &	4.8 (1.5)\\ \cline{2-11}
& \multirow{2}{*}{100} 
& 93 & 90.8 &	75.7 &	52.3	& 99.7	& 38.6 (4.4)& 78.0	& 93.9	& 3.4(0.7)\\ 
& & 91 &	90.5 &	80.4	& 47.9	& 99.5	& 40.3 (5.7)	& 83.3	& 92.8	& 4.4 (1.5) \\ 
\hline
\multirow{8}{*}{15} 
& \multirow{2}{*}{ t-VAF } 
& 99 &	91.7	& 85	& 63.1	& 100 &	40.8 (2.2) &	88 &	96.6	& 2 (0.2) \\
& & 96&	92.1&	87.5&	61.7&	100&	41.5	(2) &89.1	&96.9	&2  (0.3) \\ \cline{2-11}
& \multirow{2}{*}{10K} 
& 98	& 92.3	& 85.3	& 61.2	& 100 &	44.3	(3.5) & 87.6 &96.2	&2.4 (0.3)\\
& & 100 &	90.6	& 83.6 &59.1	&100&	41.8	(2.6) &85.8&	96.6&	2.1 (0.3) \\ \cline{2-11}
& \multirow{2}{*}{1K} 
& 93 &  94.7 &	83.5	& 61.5	& 99.9	& 42.7 (4.3) & 85.4	& 96.1	& 2.6 (0.6) \\
& & 100 & 91.8 &	85.4 &	59.1	& 100 &	43.5 (2.9) &	87.8&	96.7&	2.1 (0.3) \\ \cline{2-11}
& \multirow{2}{*}{100} 
& 92 &	81.1	& 55.2	& 48.4	& 99.4	& 37.4 (5) & 61.3	& 96.1	& 2.1 (0.3)\\ 
& & 98&	81.8	& 57.1 &46.1	&99.8	&36.5 (2.2)	&63.1	&96.2	&2.2 (0.4) \\ 
 \hline
\end{tabular}
\caption{Topological ancestry-descendant and sibling relationship reconstruction on simulated data. Results are shown for 5, 10, 15 samples given true VAFs, 10000X, 1000X, and 100X coverage data (without CNVs) obtained with localized (top row) and randomized (bottom row) sampling. All values are averaged over the number of reconstructed trees (Trees) out of 100 simulated trees. The following metrics are presented: SSNVs present in the tree (\% SSNVs), ancestor-descendant pairs of mutations in the tree (AD), ordered AD pairs (AD-Ord), correctly ordered AD pairs (AD-Corr), unordered AD pairs reconstructed as siblings (AD$\rightarrow$Sib) (with and without private mutation nodes), sibling pairs of mutations in the tree (Sib), correctly reconstructed sibling pairs (Sib-Corr), sibling pairs reconstructed as AD (Sib$\rightarrow$AD) (with and without private mutation nodes). }
\end{center}
\end{table}
\end{landscape}
}
\afterpage{
\clearpage
\begin{landscape}
\begin{table}[t!]
\begin{center}
\label{table3}
\begin{tabular}{ | l | l | l | l | l | l | l | l | l | l | }
\hline
Samples & Trees & \% SSNVs & \% AD & \% AD-Ord & \% AD-Corr & \% AD$\rightarrow$Sib (-priv) & \% Sib & \% Sib-Corr & \% Sib$\rightarrow$AD (-priv) \\ 
\hline
\multirow{2}{*}{5} 
& 95 &	96.2	& 89.1 &	14.4	& 93.7	& 20 (0.4) &	92.6 &	80.8 &	2.4(0) \\ 
&98	&97.42	&91.96	&10.74	&96.1&	0.20&	94.85	&80.46	& 2 (0.02) \\ \hline
\multirow{2}{*}{10} 
& 91 &	94.8	&77.1	&14.6	&93.5	&24.9 (0.6)	&90.9	&91.8	&1.3(0) \\ 
& 92	&94.55	&78.97	&11.95	&93.1&	0.41&	90.5 & 91.7 &	1.1 (0.01) \\ \hline
\multirow{2}{*}{15} 
& 97 &91.8	&65.9	&11.1&	92.5&	23.3 (0.4) &	85.5	&94.6&	0.7 (0)\\ 
& 94 &	92.9 &	69.6 &	10.8 &	94.5&	24.2 (0.35)	& 87.5&	94.8 &	0.6 (0.01)\\ \hline
\end{tabular}
\caption{Topological ancestry-descendant and sibling relationship reconstruction on simulated data in the presence of CNVs. Results are shown for 5, 10, 15 samples given 1000X coverage data obtained with localized (top row) and randomized (bottom row) sampling with ~80\% of SSNVs in CNV regions. All values are averaged over the number of reconstructed trees (Trees).}
\end{center}
\end{table}
\end{landscape}
}

\begin{backmatter}

\section*{Competing interests}
SB is a founding advisor of DNAnexus Inc, and member of the Scientific Advisory Boards of 23andMe and Eve Biomedical.
  
\section*{Author's contributions}
VP, RS, and SB developed the method. VP implemented the LICHeE program and simulator and drafted the manuscript. RS, SB, DK, IH contributed to the manuscript. VP, RS, SB performed the evaluation and analysis. All authors were involved in discussions.

\section*{Acknowledgements}

Authors would like to thank Arend Sidow for valuable discussions and Aaron C. Abajian for contributing to the simulation studies. \\
\noindent VP was supported by the Stanford-KAUST grant. RS and IH were also supported by Natural Sciences and Engineering Research Council of Canada (NSERC) Postdoctoral Fellowships. DK was supported by an STMicroelectronics Stanford Graduate Fellowship. This work was funded by a grant from KAUST to SB.
  
\bibliographystyle{bmc-mathphys} 
\bibliography{lichee_arxiv}   

\end{backmatter}

\section*{Appendix A --- Evaluation of Phylosub}
We ran PhyloSub [29] on the Gerlinger \textit{et al} dataset [21] using default parameters.  For each patient in the dataset we compared the top tree structures (by default, three) with the published tree. It is hard to find similarities between the topology of reported trees and the published trees. Therefore, we analyzed the performance of PhyloSub in finding the correct clusters of mutations, which is a preliminary requirement for building the cancer progression pathway. Gerlinger et. al have categorized mutation groups as shared, heterogeneous (partially shared), and private. Generally speaking, we observed that shared mutations were clustered correctly but partially shared mutations were often clustered all together as one group, which violates their presence pattern. As an example, PhyloSub reported a tree for patient EV003 with three mutation groups (see Figure 8 below). The trunk branch (colored blue) includes all shared mutations as well as one private and two partially shared mutations. Second branch (colored green) contains several partially shared mutations and the rest of private mutations with very different presence patterns, and the third branch (colored red) contains three partially shared mutations. Case EV003 has a heterogenous group with 8 mutations shared by R9 and R6 (LONRF2, RHOB, BHLHE40, CMYA5, NOD1, GCC1, SLC5A12, YLPM1), whereas PhyloSub puts two of them in the trunk branch, four in the second, and two in the third brach. Note that, as discussed in the Results section, for EV003 LICHeE was able to find all mutation groups and build the lineage matching the reported tree in Gerlinger et. al dataset. For other cases PhyloSub showed similar performance. 
We also observed that PhyloSub does not deal with private mutations properly. Often private mutations are either clustered into singleton groups or are distributed into shared or partially shared groups. For example, for six out of eight patients in at least one of the top-trees there are private mutations clustered with shared mutations and located in the trunk branch of the corresponding tree. Note that PhyloSub shows acceptable performance in analyzing samples with very few mutations and simple (chain) topology.
However, when tested on more complex multi-sample cancer datasets, the performance of LICHeE is far more superior. 

\begin{figure}[t!]
\includegraphics[scale=.3]{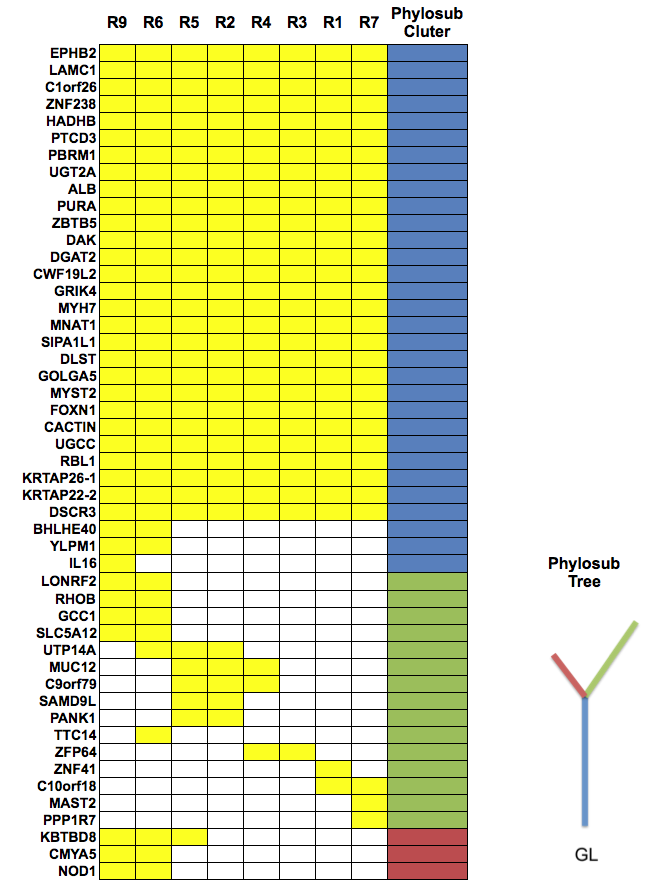}
\caption{PhyloSub results for patient EV003.}
\end{figure}

\section*{Appendix B --- Experimental Details of the ccRCC, HGSC, and Xenoengraftment Comparison}

\textbf{ccRCC.} The clear cell renal cell carcinoma (ccRCC) study by Gerlinger et. al [21] validated 602 nonsynonymous nucleotide substitutions and indels from multiple samples of 8 individuals using ultra-deep amplicon sequencing with an average depth of $>$400x. The analysis was performed on 587 out of these mutations guaranteeing a minimum coverage higher than 100x. Based on the expected sequencing platform error rate, mutations were called in a sample if their VAF was  $\geq$ 0.5\% for substitutions and 1\% for indels. In accordance with the study, we set the cutoffs to $T_{present}$ = $T_{absent}$ = 0.005 for calling SSNVs across samples for all patients except RMH004 and required at least two mutations as evidence of a node in the tree. For RMH004 we set $T_{present}$ = $T_{absent}$ = 0.01 to allow editing, observing that some mutations with the frequencies slightly higher that 0.5\% were considered absent in some samples of the study.\\

\noindent \textbf{HGSC.} In the high-grade ovarian cancer (HGSC) dataset from the study by Bashashati et. al. [27], 19 tumor samples from six patients were used to validate 340 somatic mutations using deep amplicon sequencing (with a median coverage of $>$ 5000x). When running LICHeE, we used the hard thresholds of $T_{absent}$ = 0.005 and $T_{present}$ = 0.01 for all the patients except Case5, where we used $T_{absent}$ = 0.01 and $T_{present}$ = 0.04 due to a higher level of noise in the data, and required at least three mutations as evidence of a node in the tree. For Case5 we also removed the samples g and h from consideration since they had multiple inconclusive validation results as stated in the manuscript [27]. \\

\noindent \textbf{Xenoengraftment.} The study by Eirew et. al. [30] used deep-genome and single-cell sequencing to evaluate the clonal dynamics of xenoengraftment of breast cancer tissue into immunodeficient mice. Single-cell analysis was done on passages SA501 (samples X1, X2,  and X4) and SA494 (samples T and X4). When running LICHeE on both passages, we used the hard thresholds of $T_{absent}$ = 0.03 and $T_{present}$ = 0.05  (the value 0.05 was indicated in the study's supplementary materials); a minimum cluster size of 6 for non-private mutations and 5 for private mutations; and a maximum cluster collapse distance of 0.085.

\end{document}